\documentclass[sigconf]{acmart}

\usepackage{balance}       
\usepackage{graphics}      
\usepackage[T1]{fontenc}   
\usepackage{txfonts}
\usepackage{color}
\usepackage{booktabs}
\usepackage{textcomp}
\usepackage{xspace}
\usepackage{enumitem}

\usepackage{todonotes}

\renewcommand\footnotetextcopyrightpermission[1]{} 
\newcommand{\blank}{\emph{Blank\_Screen}\xspace}
\newcommand{\abs}{\emph{Abstract\_Visualization}\xspace}
\newcommand{\eyes}{\emph{Realistic\_Visualization}\xspace}

\begin{document}
\title{Frontal Screens on Head-Mounted Displays to Increase Awareness of the HMD Users' State in Mixed Presence Collaboration}


\author{Christian Mai}

\affiliation{%
  \institution{LMU Munich}
  \postcode{43017-6221}
}
\email{christian.mai@ifi.lmu.de}

\author{Alexander Knittel}
\affiliation{
  \institution{LMU Munich}}

\author{Heinrich Hu\ss{}mann}

\affiliation{
  \institution{LMU Munich}}
\email{hussmann@ifi.lmu.de}

\renewcommand{\shortauthors}{C. Mai \& H. Hu\ss{}mann}

\begin{abstract}
In the everyday context, e.g., a household, HMD users remain a part of the social life for Non-HMD users being co-located with them.
Due to the social context situations arise that demand interaction between the HMD and the Non-HMD user.
We focus on the challenge that the Non-HMD user is not able to interpret the HMD user's state -- e.g., attentiveness; the need for assistance --, as the HMD covers the wearer's face.
We propose a front facing display attached to the HMD that supports collaboration by showing the state.
We explore the impact of abstract and realistic visualizations for such displays on collaborative performance and social presence in a within-subject user study (N=25).
We present to the Non-HMD user (1) a blank screen (baseline), (2) textual representation of the user's state and (3) a representation that looks like the HMD is see-through.
The results show positive effects for textual representation on collaborative performance and a positive effect of realistic representation on social presence.
We conclude that when developing HMDs we need to take into account the social needs of everyday life to reduce the risk of social separation in a household context.
\end{abstract}


\begin{teaserfigure}
\centering
    \includegraphics[width=\textwidth]{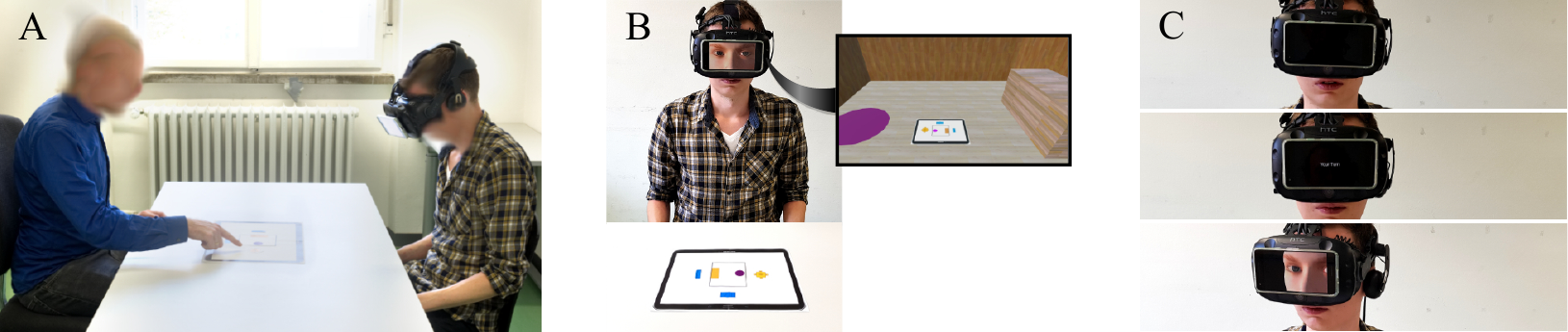}
    \captionof{figure}{In a co-located mixed-presence situation (A) the Non-HMD user has difficulty recognizing the state of the HMD user, due to the HMD covering the face. For efficient collaboration environmental awareness (shared floorplan, B) and awareness for the other is necessary (topic of this paper). We suggest to use a front facing display (B) to restore awareness about the HMD user's state. We compared three types of visualizations (C) black screen (baseline), an abstract visualization (text) and a realistic visualization animated according to the current state.}
    \label{fig:teaser}
\end{teaserfigure}

\maketitle

\section{Introduction}
The advent of consumer-grade head-mounted displays (HMDs) leads to situations in which people use HMDs outside of protected laboratory environments.
Possible examples of these emerging environments are households in which a mother is immersed in a virtual reality meeting, while other family members approach her for typical request during everyday life.
Companies offer virtual reality experiences to support the imagination of the costumer before a possible purchase decision. 
Examples are cars~\cite{audi.2016,Blackstone_18,volveSUV}, holiday offers ~\cite{marriot} or kitchen design at the point of sale ~\cite{inno}.
In both situations Non-HMD users are around the HMD user, creating a social context that did no exit in the laboratory context.
We call  (Figure \ref{fig:teaser}, A).

We focus on the Non-HMD user, who might have obligation to be present in the real world or just no interest to be in the virtual one.
In all cases, the communication with the HMD user is impaired in this \emph{co-located mixed-presence} situation for two reasons ~\cite{McGill:2015:DRO:2702123.2702382}:
(1) The Non-HMD user does not share the environment with the HMD user. The Non-HMD user does not know about the HMD user's position in the virtual world or the state the virtual character is in.
(2) The HMD covers the user's face, in particular the eyes, making it difficult for the Non-HMD user to interpret the state the HMD user is in. 
In this work we aim to challenge (2), as we argue in the following.

Gaze is used in communication to provide information, regulate interaction, express intimacy, exercise social control and facilitate services or task goals~\cite{Patterson}.
We focus on restoring regulating and monitoring mechanisms of gaze, which we call the state, visible to a Non-HMD user.
Particular examples are synchronizing turns during communication~\cite{duncan77,10.1371/journal.pone.0136905,Kendon67}, referencing an object~\cite{BOCK2008946,Prasov:2008:WGR:1378773.1378777,Symons2004WhatAY} or being aware of each other~\cite{Kendon67}.

The problem is that disturbed communication between the HMD and the Non-HMD user leads to frustration, failing in collaboration and can result in taking off the HMD.
The HMD is taken off to restore regular communication, without any visual barrier.
With the increasing use of HMDs in household environments, the result of a failed interaction could be even more dramatic.
Repeated failure can lead to an HMD user no longer being addressed in the household and thus being excluded from everyday social life.
This risk of excluding the HMD user is gaining attention in the scientific community.
Related work proposes different kind of visualization and interaction metaphors,
we discuss in more detail in the following chapters~\cite{Gugenh19,Gugenheimerface2017,Mai_18_Nordichi,Pohl.2016,Weyers2018}.

Therefore our main \textbf{research questions} is:

\begin{itemize}
    \item How to design the communication of an HMD user's state, naturally communicated by eye-gaze, towards an bystander?
\end{itemize}

To target our research question we present design categories that support the classification and design of information displays.
Further these design categories reveal open questions and motivate our system designs (Figure \ref{img:DesignSpace}).
In a within-subject user study (N=25) we explore the effects of a front facing display attached to an HMD on social presence and task performance (Figure \ref{fig:teaser}, B).
We do so by the three conditions (Figure \ref{fig:teaser}, C): (a) \blank: baseline, (b) \abs:  showing textual information about the users state on the front of the HMD and (c) \eyes: creating the illusion of looking onto a real face inside the HMD. 

In summary our \textbf{contributions} are:

\begin{itemize}
    \item Design categories for displays communicating the HMD user's state, naturally transmitted by eye-gaze, in co-located mixed-presence situations
    \item Insights on the the design of a front-facing information display attached to an HMD for the design category \emph{Visualization}
\end{itemize}

\section{Designing for the Visual Barrier}
\label{sec:design}
\begin{figure}
    \includegraphics[trim = 0 36 0 0, clip]{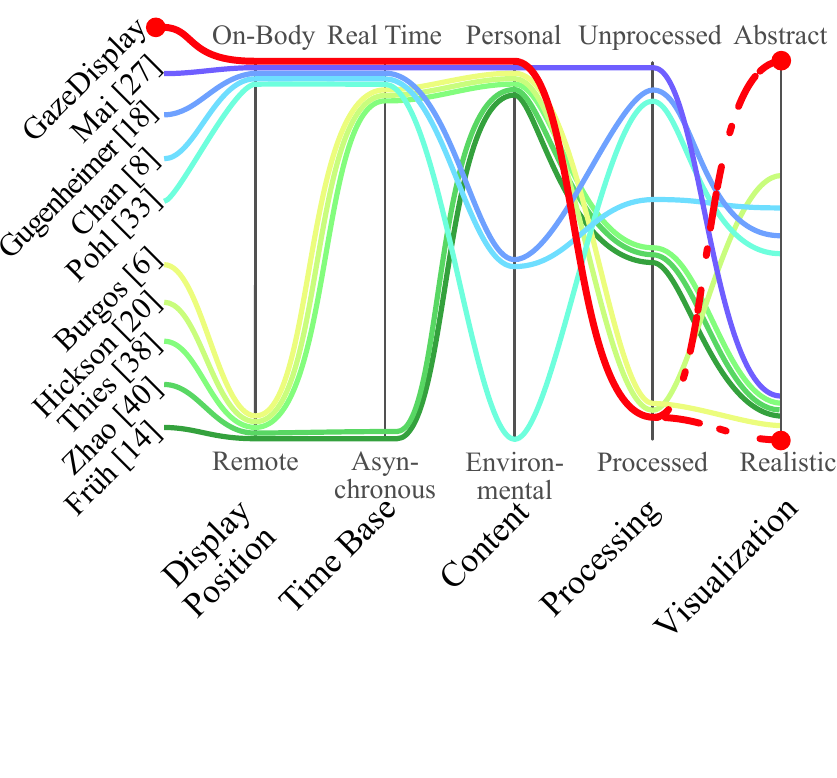}
    \caption{Design categories for communication of the users state, naturally expressed by eye-gaze, in mixed-presence scenarios. Red shows the design of our proposed system and conditions used in our user study. The upper ones are explicitly designed to overcome the visual barrier in co-location. The lower ones use a monitor to showcase the systems ability. It becomes visible that the On-Body visualizations mainly focused on a visualization of the users state somewhere in between abstract and natural, while the remote ones focused on realism. We close the gap with our approach.}
    \label{img:DesignSpace}
\end{figure}


To delimit the scope of our work from related work on transmitting information expressed by the HMD user's face, we derive \textbf{design categories}.
The resulting categories, with the according related work sorted into, are shown in Figure~\ref{img:DesignSpace} and will be used to derive our concept. 

As argued in the introduction the HMD user acts in two environments, the virtual and the physical.
Therefore, information associated with the user may be either part of the virtual environment and the virtual character, or part of the real environment and the user's personal information.
Although our design categories might be applicable for both, in this work we only look into supporting the transmission of personal information that is not visible for the bystander due to the HMD covering the face.
Other personal information might be physiological -- e.g., heart rate -- or information about the virtual character the user is controlling -- e.g., position in the virtual environment or health of the character --.

In the following we will briefly explain the categories and their characteristics, accompanied by the examples from related work.

\emph{Display Position} describes the visualization display's location. The scale is binary with the characteristics \emph{remote} and \emph{on the body} of the HMD user. \emph{Remote} presentation would be the display of the user's state on a wall mounted monitor, while the user walks through the room~\cite{Frueh:2017:HRV:3084363.3085083}. 
A display on the face of the user belongs to the characteristic \emph{On-Body}~\cite{Gugenheimerface2017}.
When designing the display position one should keep in mind that there are subscales of this category that influence for example the readability. 
When the display is positioned on the back of the HMD user, the Non-HMD user will not be able to see information on it when they are facing each other.

\emph{Time Base} of a visualization is either \emph{asynchronous} or \emph{real-time}. 
\emph{Asynchronous} visualization is apparent, when the Non-HMD user perceives a timely disconnection between the HMD users actions and the presentation his/her state. 
A \emph{real-time} visualization shows the state of the user synchronized with their actions. 
A \emph{Real-Time} example is the animation of eyes on the front of the HMD by eye-tracking data~\cite{Chan:2017}.  
\emph{Asynchronous} presentation is a video with an overlay of the user's face created in post production~\cite{Burgos2015}. 

\emph{Content} communicates information about the user's state either by showing the virtual \emph{environment} or \emph{personal} information. 
Chan and colleagues~\cite{Chan:2017} communicate that the HMD user is focusing his/her attention on the virtual world by showing the virtual environment on the screen. 
Not showing the virtual world would indicate that the HMD user is attentive to the real world.
The same could be transmitted by showing eyes on front of the HMD. 
Open eyes could be interpreted as being attentive to the real world and closed eyes as inattentiveness.

\emph{Processing} level of the data is described by the processing of the input data. Presenting a video stream of an eye-tracking camera we define as minimally processed data.
Processed data implies an intended manipulation of the data, leading to an abstraction of the original data.
An example would be the analysis of the eye tracking information and derivation of the user's emotional state~\cite{Frueh:2017:HRV:3084363.3085083}.

\emph{Visualization} of the transmitted information means either showing an abstract or realistic visualization of the data. 
The characteristics of this category can be expressed by a continuous line between symbolic on the one end and naturalistic on the other.
A visual abstraction, e.g., the substitution of the users real face by a comic-like avatar or even a number~\cite{Chan:2017,Frueh:2017:HRV:3084363.3085083,Gugenheimerface2017}.
Showing a video of a face on would have a less realistic appearance then a real life avatar matching the HMD users covered face with fitting perspective for the Non-HMD user,creating the illusion of looking through the HMDs case\cite{Mai.2017}.

\subsubsection{Related work} A number of applications aimed on making the HMD covering the face optically disappear. We divide them into work on tracking and visual effects, as well es particular designing for the co-locale mixed-presence situation.

The presentation of the HMD user in a video clip with the 3D overlay of the user's actual face is an example of an asynchronous visualization on a remote screen~\cite{Thies2017FaceVRRF}. 
It is possible to create the overlay in a real-time video stream~\cite{Burgos2015,Frueh:2017:HRV:3084363.3085083} and even the animation of facial expression of the superimposed avatar in the video stream is feasible~\cite{Burgos2015,Frueh:2017:HRV:3084363.3085083} by recognition of facial expression~\cite{Burgos2015,Thies2017FaceVRRF,2018arXiv180708772Z} and emotional state~\cite{HicksonDSKE17}.
These advances give us good indications that the detection of the users state is technically feasible.
Therefore our work does not dive into the technical improvement of the detection or simplification of the systems.
Our work focuses on the system design that might base on such technology.
The existing solutions did not yet consider the opportunities that come with the co-local situation the HMD and the Non-HMD user are acting in.

\emph{FaceDisplay}s enable the bystander to look inside the virtual world on displays mounted around the HMD~\cite{Mai.2017,Gugenheimerface2017,Pohl.2016}.
An advantage of such an information display is the transportability~\cite{Gugenheimerface2017}.
Gugenheimer and colleagues~\cite{Gugenheimerface2017} introduced FaceDisplay as an interactive system fostering the asymmetry of capabilities of an HMD and Non-HMD user.
In contrast to Gugenheimer, we do not aim on leveraging the asymmetry between the co-locaties, but on enabling a collaboration with symmetric availability of information as it is needed for successful collaboration~\cite{Gutwin2001TheIO}.
Further we want to transmit personal information of the users state, e.g., attentiveness, and not information about the virtual environment and characters as in the playful experiences of Gugenheimer.

Chan and Minamizawa propose to use a display on the front of the HMD in combination with an eye-tracker~\cite{Chan:2017}. 
The eye-tracking information is used to animate 2D eyes in a cartoon-like design.
The HMD user's focus of attention, either on the virtual or the real world, is indicated by changing the background presented on the display, which helped the Non-HMD user.
However their results are based on a preliminary focus group and due to the cartoonish look, does not take into account the degree of abstraction as described by the category \emph{Visualization} (Figure \ref{img:DesignSpace}).

Mai and colleagues create the illusion that one can look through the HMD onto the wearers face was~\cite{Mai.2017}. 
They did not evaluate it, nor was the purpose to communicate any information.

Many solutions have been proposed, but there is a gap in research between the approaches focusing on video streams and the once on designing solutions as shown in Figure \ref{img:DesignSpace}.
Video based approaches use \emph{remote} screens to communicate interaction and designing solutions use for displays \emph{on the body}.
Further technical solutions try to achieve a very \emph{realistic} visualization and the design based solutions focus on more or less \emph{abstract} visualization.
An abstract visualization was not used yet in both domains. 
The technical approaches have numbers, e.g., describing the emotional state, but represent them with avatars.
In our work we therefore explore the gap between realistic and abstract visualization of the user state.

\section{Concept}
\label{sec:concept}
In the following we present our system and study design.
We motivate it by introducing the role of gaze in human collaboration first.

\subsection{Establishing a Common Ground in Collaboration}
\label{sec:commonGround}
As described in the introduction, Patterson divides gaze and its role in human communication into five categories ~\cite{Patterson}.
On a more abstract level, the use of gaze can be divided into the communication and regulation of social relationships as well as regulating and monitoring mechanisms during an interaction. 
In our user study, we focus on the regulating and monitoring mechanisms of eye-gaze.
These functions of gaze have immediate utility in collaboration and help us to quantify the effects of eye-gaze.

Maintaining awareness for the other person and their situation is crucial for efficient collaboration.
Awareness in successful collaboration not only depends on spoken words.
Awareness also depends on signals from the environment, the shared workspace, and (gaze) cues by the other collaborator~\cite{Gutwin2001TheIO}.
To maintain awareness, three sequential steps can be depicted~\cite{Endsley95}:

Firstly, collaborators need to \emph{perceive} the current state. 
Secondly, they have to \emph{comprehend} the current state. 
Finally, collaborators need to \emph{project} the current state into the future in order to make a conscious decision about the next steps to undertake, e.g., reacting to the situation.
For situation awareness it is necessary to maintain a \emph{common ground}~\cite{clark1991grounding}, which should be supported by a communication system~\cite{Endsley95}. 


\subsection{System Design and User Study}
Since our work focuses on the awareness of the other person, we introduce a shared surface in our system design, to provide a common ground~\cite{Mai_Mum18}.

\subsubsection{System Design}

Looking at the design categories, one can see that realistic representations of a users face behind the HMD are dominantly used on remote screens (Figure~\ref{img:DesignSpace}), lower five). 
The purpose of the related work, is focused on technical applications that target on creating video streams for the web.
In contrast our work focuses on designing for the co-located mixed-presence situation. 
Related work on this till now used abstract visualization of the user face to support the collaboration (Figure~\ref{img:DesignSpace}), upper four).
In our work we want to bridge the gap and explore the full range of the category \emph{visualization}.
Therefore we propose to use a strong abstraction of the user's state (\abs) and compare it to a realistic visualization  (\eyes).
We choose text as the highest form of possible abstraction. 
There might be other possibilities like using colors to code the user's state, which might intoduce a bias to the user study. 
A strong abstraction would need additional training of the user and, depending on the type of abstraction, is subject to cultural factors and trends.


We mount the display on front of the HMD (\emph{OnBody}) as the face is a natural location to look at for real-world bystanders when searching for information about the HMD user.
The input for the system is based on data processed in real time, that will be simulated in our user study in order to prevent errors in the system to influence the outcome of the study.


For the \eyes, we follow the idea of Mai to create the illusion of looking through the HMD's case onto the user's face ~\cite{Mai.2017}.
However, human beings are very sensitive when being confronted with a life-like anthropomorphic avatar, which has its negative extreme in the Uncanny Valley effect~\cite{Mori12}.
This effect creates a sense of unease in the observer, when a model looks a lot like a real human, but not exactly as a real human.
In an iterative design process, we optimized our visualization to avoid an Uncanny Valley effect for the Non-HMD user.
We took into account the lighting conditions, realistic shading of the Avatar, texturing, animated the face and eyes ~\cite{Lee:2002:EA:566654.566629} to show realistic movements (Figure ~\ref{img:gazeCues}).

As a baseline, we chose the \blank condition (Figure~\ref{fig:teaser}, C top), not showing anything, which is the current state of the art.

\subsubsection{Gaze Cues as Stimuli}
\label{sec:GazeCues}
Based on the deliberations on awareness in collaboration above we decided to use four exemplary gaze cues, from the set of regulating and monitoring cues, as stimuli in the user study.

We do not aim on quantifying social cues of gaze, as they are challenging to imitate and the variety of possible patterns varies. 
Furthermore, the interpretation of social cues also includes, among others, the social relation to the other and there is a natural chance of misinterpreting the meaning~\cite{kleinke86}.
Nevertheless, we expect the existence of a vivid face to influence the Non-HMD user's feeling of social presence towards the HMD user.
We follow Bioccas definition of social presence as a multidimensional construct comprising perception of co-presence, psychological involvement and behavioral engagement~\cite{Biocca01}.

In the user study we embedded the gaze cues in the communication flow that was directed by the task. 
Other modalities like pointing with a finger might be used, however they demand to be not occupied elseway, e.g., in the VR and we explicitly look into communication of states that are naturally communicated by gaze.
We do not simulate this in the user study.

\begin{figure}[t]
\includegraphics[width=\columnwidth]{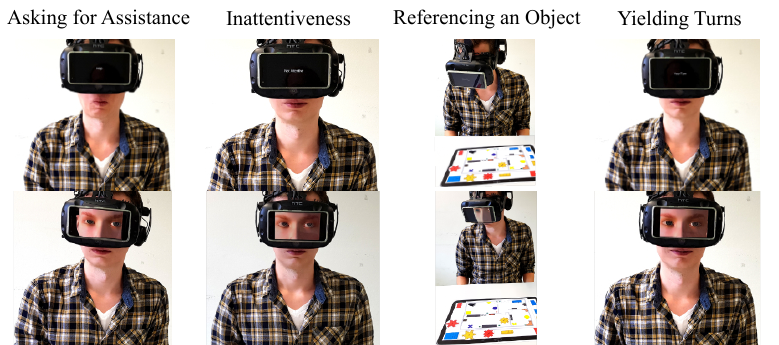}
    \vspace{-0.5cm}
    \caption{The gaze cues used as stimuli in the user study: \abs (top) and \eyes (bottom).}
    \label{img:gazeCues}
\end{figure}

We motivate the chosen gaze cues and the according implementation in the user study for the \abs and \eyes condition (Figure ~\ref{img:gazeCues}):

\emph{Asking for Assistance} -- 
A common ground needs to be established for successful collaboration and reestablished by the communication partners when broken down. 
A communication system needs to support this process, e.g., by providing information about the others state through gaze.
People are very sensitive in recognizing the gaze onto their face~\cite{10.2307/1419779}. 
The presented text was \textit{``Help''}.
The gaze pattern to ask for assistance is a glance at the other when an action is expected ~\cite{Lee07}.

\emph{Showing Inattentiveness} -- 
Not being attentive towards the common task can have many reasons.
The HMD user might be involved in a task in virtual reality while the Non-HMD user might interpret the HMD user's actions as being related to the common task.
Having the two spaces in one physical place is unique to the co-located mixed-presence situation.
Chan changed the front-screen visualization by showing the virtual environment when the HMD user focused on the VR ~\cite{Chan:2017}.
We argue that this abstract visualization takes away naturalness in the interaction leading to a decrease in social awareness. 
The presented text was \textit{``Not Attentive''}.
The pattern we chose for displaying inattentiveness is gaze aversion ~\cite{Lee:2002:EA:566654.566629,Kendon67}.

\emph{Referencing an Object} -- 
Creating joint attention is crucial in tasks when two persons share a workspace~\cite{clark1991grounding}.
Collaborators are very likely to use words like ``this'' or ``these'' to refer to the object of interest. 
In order to understand what the speaking person is referring to, additional information is needed.
Often a reference to a specific object goes together with a gaze towards the particular object.
There might be situations in which one would use the hands, however this demands that the hands are not occupied with a different task.
Gaze showed to have a high influence on the communication flow ~\cite{Kendon67,Prasov:2008:WGR:1378773.1378777}. 
Processing the information which object another person is looking at is quite complex.
The accuracy of gaze perception towards an object is still very high ~\cite{Symons2004WhatAY,BOCK2008946}.
The reference to a particular item was made in both conditions in conjunction with the HMD user turning the head towards the object. 
As text, \textit{``This One''} was presented.
The chosen gaze pattern for referencing an object with the eyes, was a gaze towards the target.

\emph{Yielding Turns} -- 
Yielding turns describes the turn allocation process between speaker and listener and might hold for cross-cultural studies~\cite{Stivers10587}.
A single cue is not enough for turn allocation.
Most of the time it is a combination of visual cues from the gaze and verbal commands by the speaker ~\cite{10.1371/journal.pone.0136905,duncan77}.
Often, ``transition ready states'' are communicated by a gaze of the speaker to the other person and the other person begins to speak with averted gaze~\cite{10.1371/journal.pone.0136905, Duncan72}.
Errors in the transition reduce the conversational flow and therefore influence the quality of the conversation or collaboration.
The presented text was \textit{``Your Turn''}.
The implemented gaze pattern was a focus with the eyes on the Non-HMD users face and a slight animation of the eyes looking at different parts of the face.

\emph{Idle-State} -- 
In the \eyes condition an \emph{Idle-State} was implement, when no gaze cue was activated.
The eyes looked straight ahead, being focused on a target about one meter in front of the HMD user.
In random time intervals, blinks and eye brow twitches were animated and the eyes showed short randomly distributed saccades.
The HMD user pretended to move through the virtual environment.
The \abs condition showed a blank screen when no cue was presented.

\section{User Study}
The experiment followed a within-subject design, using the level of abstraction from the category \emph{representation} as independent variable with the three levels \blank, \abs and \eyes (Section \ref{sec:concept}).
The participants took the role of the Non-HMD user.
The examiner took the role of the HMD user, performing a wizard-of-Oz experiment.
We did not use eye-tracking and face-tracking to detect the users' state, as this would have created high effort for the implementation, with a high probability to introduce failures in the user study like erroneous detection's or latency.

\subsection{Participants}
We recruited 25 participants mainly from an academic background, with a mean age of 26 (SD = 2.8;  52\% female) through paper displays, online social media and e-mail. 96\% had prior experience with virtual reality systems. Participants were compensated with 10 Euros or study credits. 

\subsection{Apparatus}
The VR system consists of an HTC Vive HMD and a VR-ready computer with Nvidia GTX 1080, IntelCore i7 6600k and 16GB RAM. 
For the front-facing display, a Google Pixel 2 Smartphone was used, with a Qualcomm Snapdragon 835 Processor, 4GB RAM and an AMOLED Display (1920 x 1080) measuring 5 inches in diameter. The front camera of the smartphone has 8 megapixels and 30 fps, used for tracking the Non-HMD user's face. 

The smartphone rendered a Unity3D scene showing either a 3D model of human eyes and face, text or a blank screen. 
For the 3D model, a body model from \emph{DAZ3D}\footnote{\url{https://www.daz3d.com}} was used. The library \emph{Random Eyes}\footnote{\url{http://crazyminnowstudio.com/unity-3d/lip-sync-salsa}} provides animations for natural blinking, eyebrow twitching and the animation of the gaze direction.
The text was designed to be smaller then a interpupilar distance of 60mm with the font-size being 24pt.

To track the interlocutor's position, the API \emph{Mobile Vision} v. 11.0\footnote{\url{https://developers.google.com/vision}} was used. This API allows the system to use the front camera of the smartphone to track the interlocutor's position in 2D and project it into the 3D scene at a given distance.

To give the examiner of the study the ability to control the cues during the task, another smartphone acts as a remote control, connecting to the system via bluetooth.

\begin{figure}[t]
    \includegraphics[width=\columnwidth]{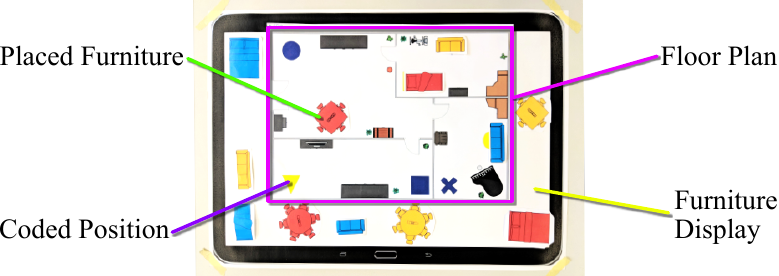}
    \caption{The shared surface was a paper prototype of a tablet, running a software that enables to furnish an apartment. Available furniture was arranged around the floorplan of the apartment. The floorplan enabled shared environmental awareness for the Non-HMD and the HMD user.}
    \label{img:floorplan}
\end{figure}

\subsection{Task}
The goal of the task was to foster the need to monitor the HMD user and create situations that require the support of gaze cues.
The participant was asked to take the role of a salesperson in a furniture store. 
The HMD user, in our case the examiner, played a customer who wished to furnish his new apartment.
The floor plan (Figure \ref{img:floorplan}) was visible for both in the same physical position, providing a common ground regarding environmental awareness (Figure \ref{fig:teaser}, B).
The participants did not know that the examiner was able to see them and the floor plan via a webcam stream.
We decided on that as our system aims on separating the HMD user and further the integration of pictures from the real world often collides with the virtual environment.
Further we wanted to avoid that user speak to the webcam, as they know that they are seen through it.
A paper prototype was used to simulate a software running on a tablet (Figure \ref{img:floorplan}). 
The participant should now ask the HMD user which furniture items he would like to place on marked spots throughout the apartment and put it on the respective spot on the floor plan.
The action was repeated until all empty spots were covered. 

\begin{table}[b]
\centering
\begin{tabular}{|p{0.16\columnwidth}|p{0.36\columnwidth}|p{0.36\columnwidth}|}
\hline
\textbf{Cue}  & \textbf{Correct Reaction}                                                                 & \textbf{Error}                                                                      \\ \hline
Asking for Assistance        & Participant offers help or implies that something is wrong.                       & Participant does not react correctly within 10 seconds.                              \\ \hline
Inatten- tiveness & Participant follows up with customer (e.g. asks how he likes the furniture item). & Participant does not react correctly within 10 seconds.                              \\ \hline
Refer- encing an Object     & Participant touches or mentions correct item.                                     & Participant touches wrong object or asks which object the examiner means. \\ \hline
Yielding Turns      & Participant responds in some way.                                                 & Participant does not react correctly within 10 seconds.                              \\ \hline
\end{tabular}
\caption{Definitions for correct reaction and false reactions to the four cues respectively.}\label{tab:measures}
\end{table}

The spots had four different shapes, coding the activation for one of the four different gaze cues for the examiner (Figure \ref{img:floorplan}). 
We describe the codes and the behavior in the following:
At the exact time, the participant placed a furniture item on a circle, the gaze cue communicating \emph{Inattentiveness} was activated. 
The \emph{Inattentiveness} gaze cue indicated the focus of the user's attention on the virtual world. 
When an item was placed on a triangle, the examiner activated the cue \emph{Asking for Assistance}, indicating that he was in need of assistance. 
\emph{Referencing an Object} was triggered, when the participant asked the examiner about which furniture item should be placed on a rectangle. 
The examiner then directs gaze at an item in one of the corners of the tablet.
Orally he mentions the type of the furniture item but not the color and size (e.g. ``this bed''). 
Since the objects in the four corners were of the same type, the participant was now required to understand the reference using the presented gaze cue. 
Furthermore, as soon as the participant finished placing a furniture item on a cross, the HMD user activates the cue \emph{yielding turns} without using any verbal commands, expecting the participant to continue with the task.
There were eight marked spots on each floor plan in total, with each of the four cues being activated two times within each of the three conditions.
In accordance with the cue, a particular reaction from the participant was expected, as described in Table \ref{tab:measures}.


\subsection{Measures}

\subsubsection{Behavioral Measures}
We define our measures according to Endsley~\cite{Endsley95}:

\begin{enumerate}
\item \emph{Recognition Time:}
Time between appearance of the stimulus on the display and participant's gaze towards the screen
\item  \emph{Reaction Time:}
Time between appearance of the stimulus and the correct response of the participant
\item \emph{Error Rate:}
Number of incorrect reactions to a stimulus. 
The definitions of correct reaction and possible errors are summarized in Table \ref{tab:measures}
\end{enumerate}

\subsubsection{Subjective Measures}

We measured social presence with three Likert-type scales, derived from work of Biocca and Harms~\cite{cogprints6743} and Poeschl and Doering~\cite{Poeschl15}.
The participants reported on their perceived attentional engagement, perceived comprehension and impression of interaction possibilities.

\emph{Semi-structured interviews} followed the leading questions about general experience, conversational flow and uncertainties during task execution, in particular with the examiner.

A \emph{Forced Choice} task asked the participants to rank the visualizations regarding their subjective positive attitude, the ability to support the communication, the effectiveness for the task and their general preference.

\subsection{Procedure}
Participants were welcomed, informed that they will be video recorded and filled in a consent form.
The participant and the examiner were seated opposite of one another (Figure \ref{fig:teaser}, left).
We told participant about the real world furnishing scenario as described in the task section.
By the help of a demonstration scenario, the participant was made to believe that the examiner saw a real-time resemblance of the floor plan (Figure \ref{img:floorplan}) in the virtual world that matched the physical counterpart (Figure \ref{fig:teaser}, B). 
After the demonstration, the task began according to the description above. 
The order of visualizations on the HMD (three groups) and the floorplan (four groups) were randomized.
After each visualization the participants filled out the social presence questionnaire and gave the semi-structured interview.
After the three visualizations participants conducted the forced choice task.

\section{Results}
In the following section, we begin with the report of the quantitative measures, followed by the qualitative measures.

\subsection{Results for Quantitative Measures}
The results in the following are gained as explained above.

\subsubsection{Recognition Time}

\begin{figure}[b]
    \includegraphics[width=\columnwidth]{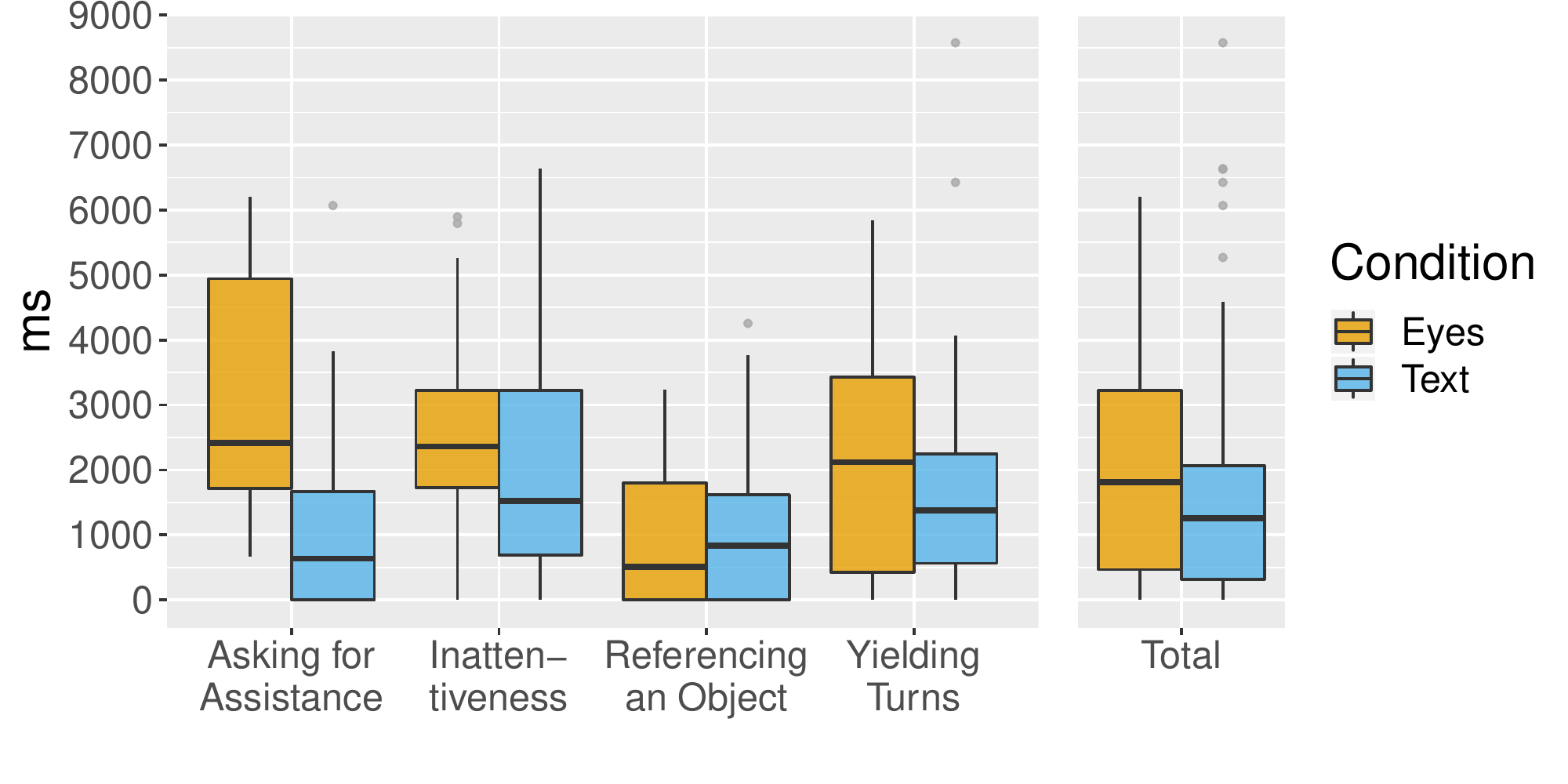}
    \caption{Recognition times for the gaze cues. Time between presenting the gaze cue and first glance to the HMD by the participant.}
    \label{fig:rectime}
\end{figure}

Considering all four cues, the recognition times averaged at 2087 ms for the \eyes condition (0 ms to 6208 ms). 
In the \abs condition, subjects recognized a cue on average in 1567 ms (0 ms to 8573 ms). 
For the \blank condition this measure was not taken since there is nothing to be recognized on a blank screen.

The results regarding the recognition time of each cue are reported in Figure \ref{fig:rectime}. 
Analyzing the data utilizing a t-test with an Alpha level of 5\% revealed that the differences in time were not significant for any of the cues. In detail, the results for the conditions are: \emph{Asking for Assistance} (t(6.042) = 1.9816, p = 0.1), \emph{Inattentive} (t(51.506) = 0.79654, p = 0.429), \emph{Reference} (t(46.115) = -0.26874, p = 0.7893) and \emph{Yielding Turns} (t(70.937) = 1.06 , p = 0.2927).

The cue \emph{asking for assistance} was recognized N=6 times out of the possible 50 in the \eyes condition and N=28 times in the \abs condition. 
\emph{Inattentive} with the \eyes condition was recognized by N=29 participants, while   N=27 recognized the cue in the \abs condition. 
The cue \emph{Reference} was recognized N=22 times using the \eyes condition and N=30 times using the \abs condition. 
The cue \emph{Yielding Turns} was recognized N=36 for \eyes and N=40 for \emph{Abstract\_Visua\-lization}. 
.

\begin{figure}[b]
    \includegraphics[width=\columnwidth]{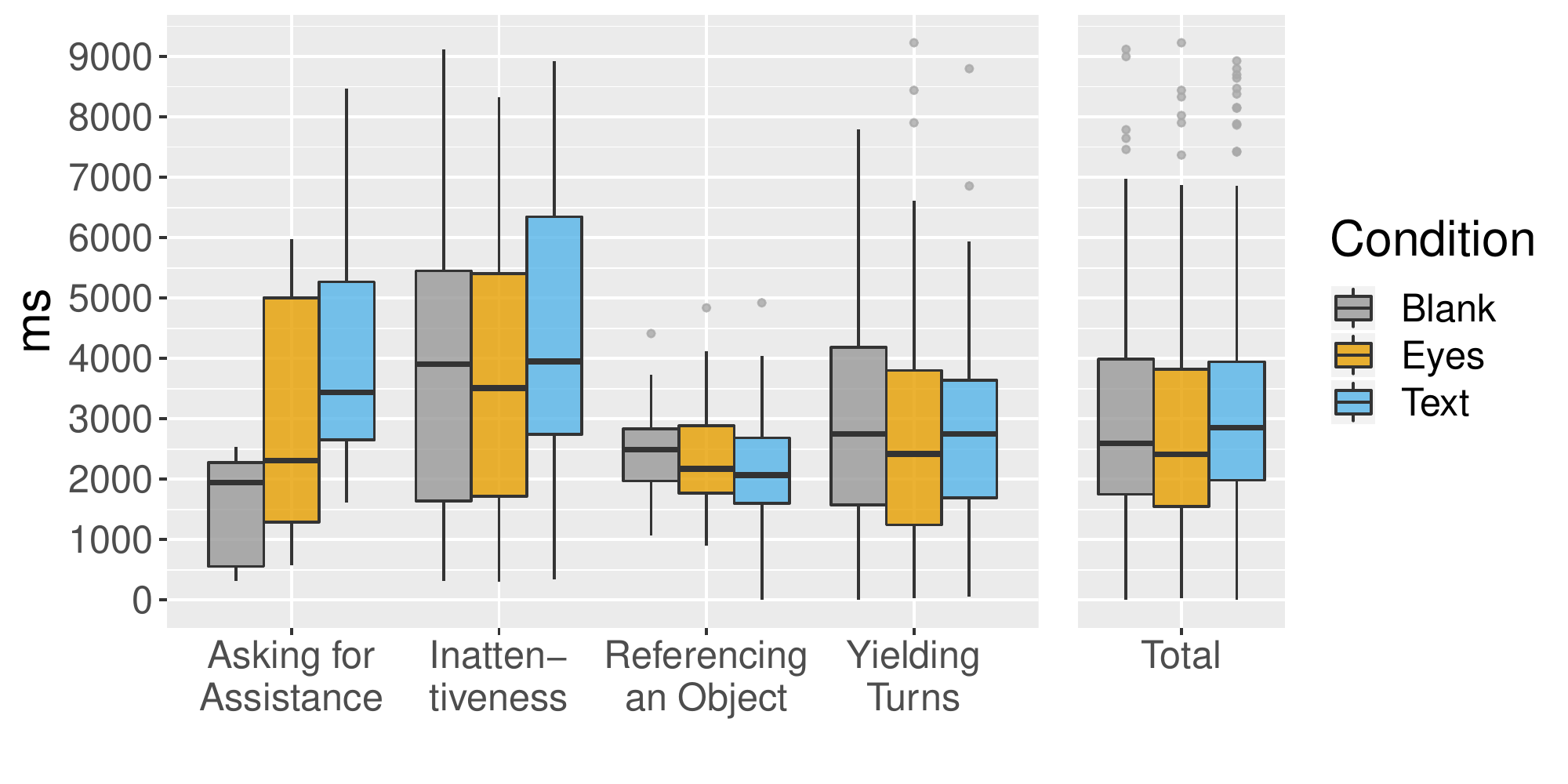}
    \caption{Reaction times defined as the time between the occurrence of a cue and correct reaction by the Non-HMD user.}
    \label{fig:reactiontime}
\end{figure}

\subsubsection{Reaction Time}

The results of all four cues on reaction time are depicted in Figure \ref{fig:reactiontime}.
Results for reaction time were compared using an ANOVA. Sphericity was tested by Mauchly's test and did not reveal any abnormalities.
Merely the \emph{Asking for Assistance} cue shows statistical significance (F(2,38) = 5.084, p=0.011) with subjects reacting significantly more slowly in the \emph{Text} condition than in the condition with no augmentation (\blank). 

\subsubsection{Error Rate}
\begin{figure}[b]
    \includegraphics[width=\columnwidth]{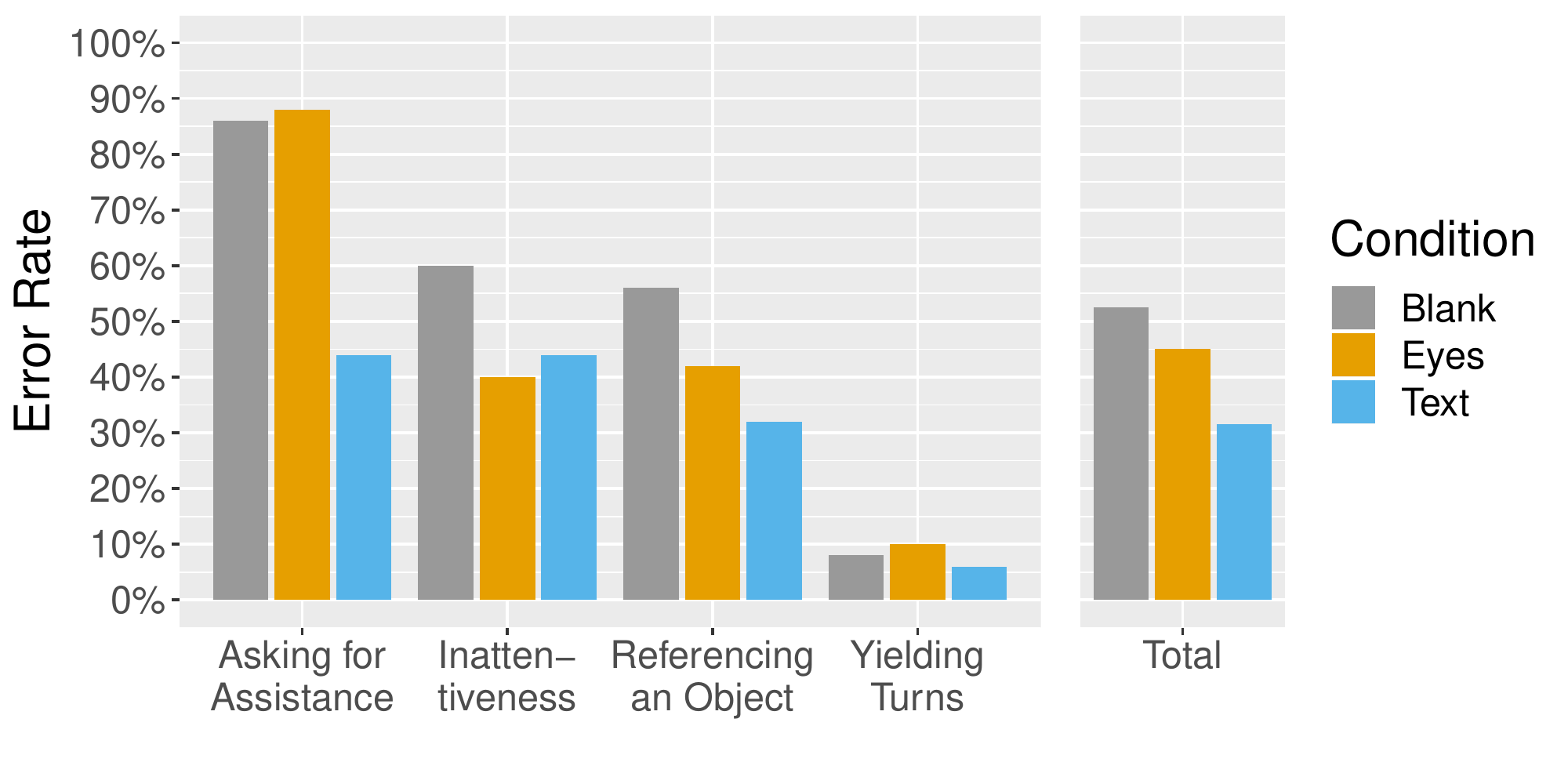}
    \caption{Error rates for each gaze cue (left) and the overall error (right).}
    \label{fig:error}
\end{figure}

Because of the binary nature of the data, we used a Chi-Square-Test in order to test the error measurements statistical significance. 
Interestingly, we could find significant differences between the \abs and \eyes conditions when comparing the overall error rates between ($\chi^2$(1)= 7.16; p= .01), as well as for \abs and \blank ($\chi^2$(1)= 17.252; p= .001).

The total error rate across all cues was 52.5\% for the \blank condition, 45\% for the \eyes condition, and 31.5\% for the \abs condition (Figure \ref{fig:error}).
When the gaze cue communicated the need for \emph{assistance}, 86\% of the subjects did not react to the cue correctly when exposed to the \blank condition and 88\% did not react correctly in the \eyes condition, while 44\% reacted incorrectly in the \abs condition.
The \emph{Inattentiveness} cue caused an error rate of 60\% for \blank, 44\% for \abs, and 40\% for \eyes.
The \emph{reference} cue showed an error rate of 56\% for \blank, 32\% for \abs and 42\% for \eyes.
The \emph{response} cue showed an error rate of 8\% for \blank, 6\% for \abs, and 10\% for \eyes.

All other comparisons did not show a statistical significance.
The results from the gaze cue referencing an object were close to significance ($\chi^2$(1)= 5.92, p=0.052).

\subsection{Qualitative Data}

\subsubsection{Questionnaire}
\begin{figure}[t]
    \includegraphics[width=\columnwidth]{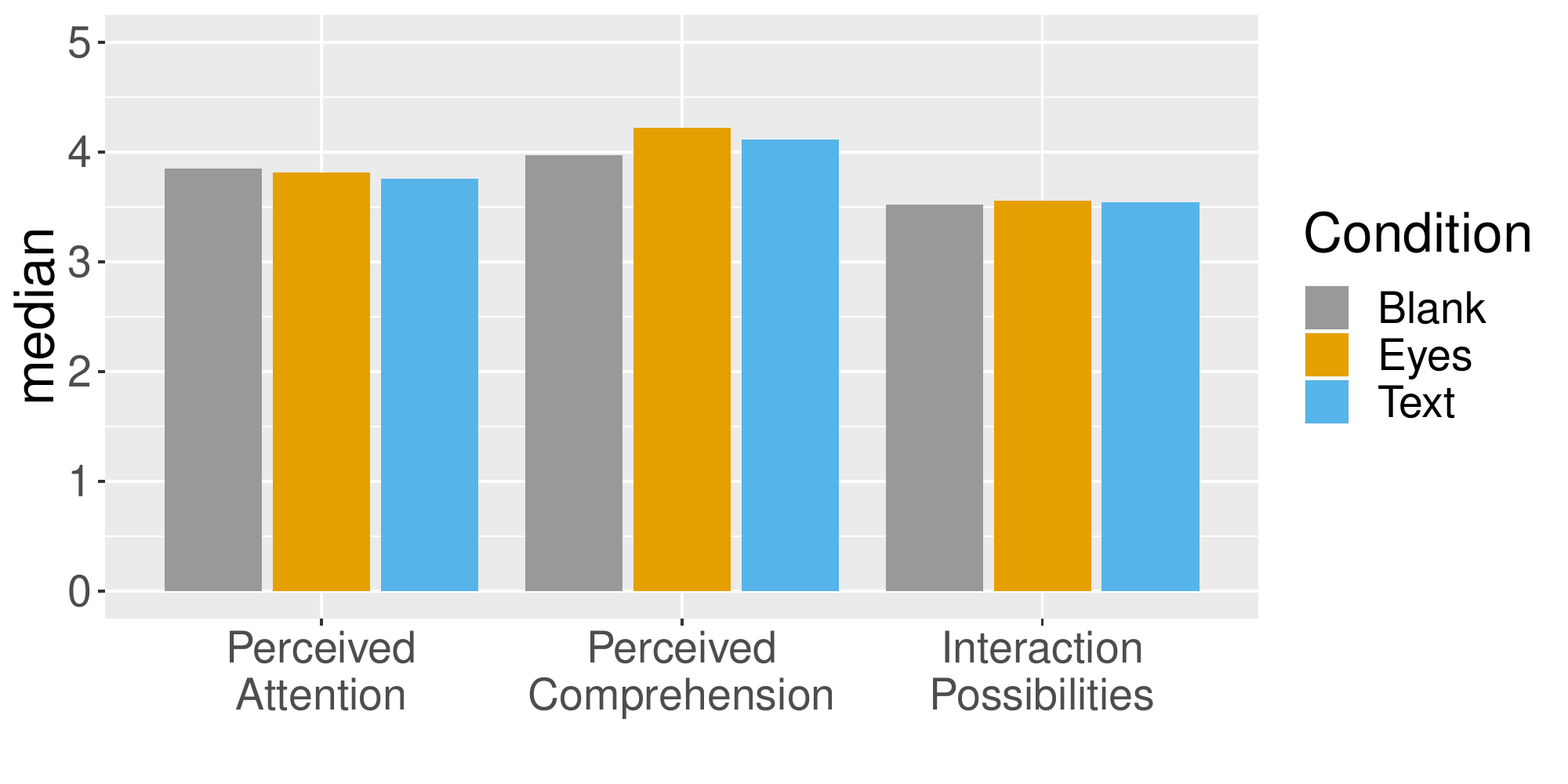}    
    \caption{The results for the social presence scales.}
    \label{fig:questionnaire}
\end{figure}

The social presence questionnaires did not show differences between the conditions.
The results for all scales show a positive tendency.
The median results are shown in Figure \ref{fig:questionnaire}.

\subsubsection{Interviews and forced choice}

The visualization of a see-through HMD split the participants into two groups either being exited or feeling uncertain.
\eyes create the feeling of being with somebody in the room (N=3) and speaking to an actual human (N=7).
While on the other hand 8 participants did not feel like talking to a human and described the augmentation as being creepy. 
6 participants reported statements related to the Uncanny Valley Effect.
The other conditions were reported to be artificial or robot-like, creating a feeling of disconnection, either in the task or personal level (\blank = 6 and \abs = 5).
Only 11 participants for condition \abs, but 19 for the \eyes condition reported to have experienced a fluent communication. 
An additional 16 participants for the \blank conditions described the communication as fluent.
8 participants reported that some cues of the \eyes were unclear. 
11 participants missed feedback while not perceiving information about the user in the \blank condition but rated the blank screen as the most effective form to collaborate (N=14).
The \abs condition was described by 4 participants as following instructions, not collaborating with a human.
It was helpful for the task when there was no need to concentrate on an opponent in the \blank condition (N=12).
Further, they reported a negative impact on the social relationship to the HMD user (N=4).

\begin{figure}[t]
    \includegraphics[width=\columnwidth]{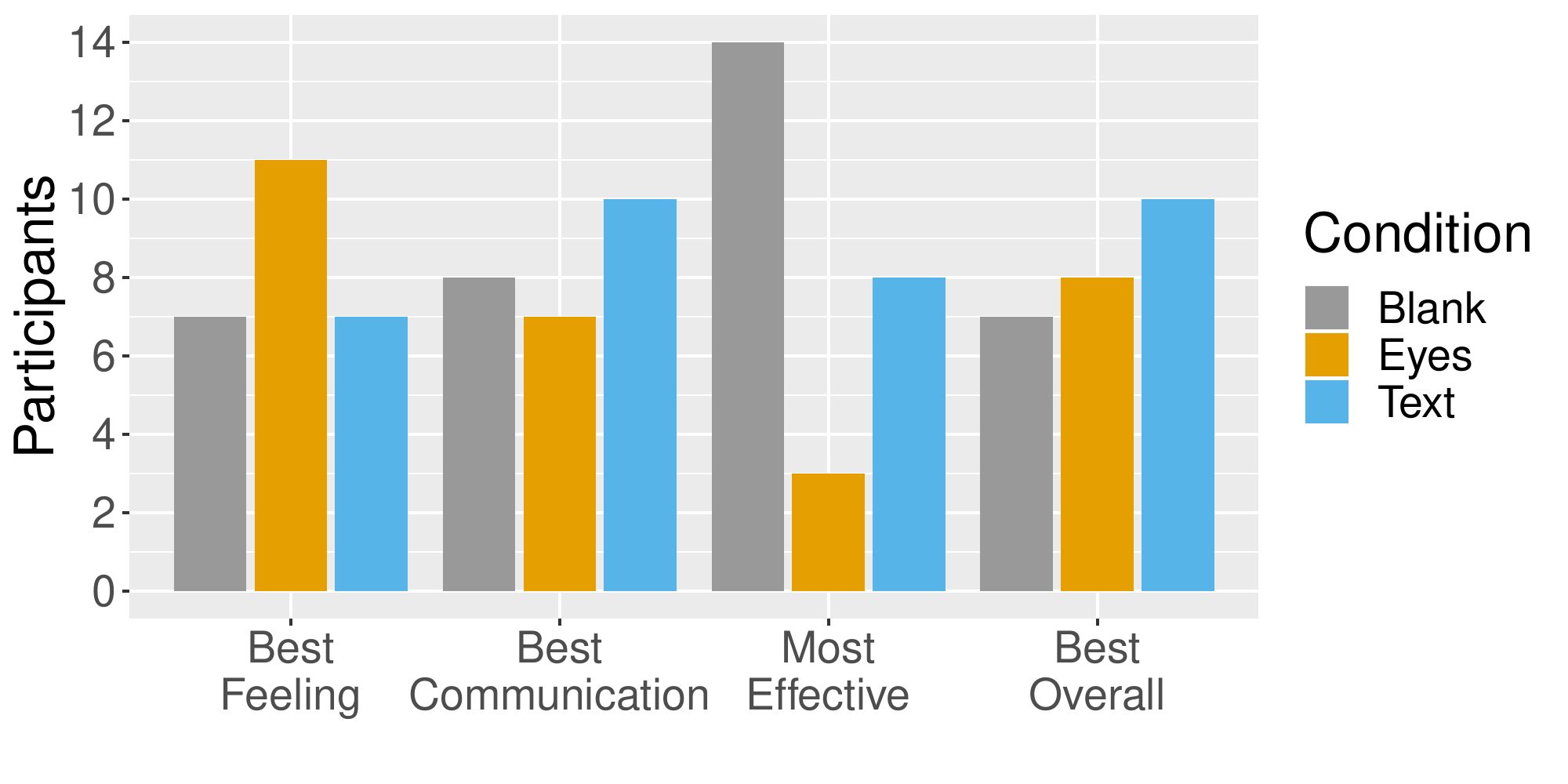}
    \caption{Sums of first choices in a forced ranking of the conditions.}
    \label{fig:subrating}
\end{figure}

After all three tasks were conducted, the participants performed the forced choice ratings.
The results are shown in Figure \ref{fig:subrating}.
Surprisingly, \abs was rated to support communication the most  (N=14) and \eyes generated the best general subjective experience.

\subsection{Limitations}
Although we took great care in the implementation of the \eyes condition, technical limitations had an effect on the experience.
When the HMD user was referencing to an object, in some cases the user was not detected by the face detection anymore. 
Therefore the animation of the depth effect freezed on the last tracked position.
The freeze led to a perspective miss-match between the displayed and the real face.
When the HMD user and the Non-HMD user changed their relative position too fast, the rendering showed latency.
The latency led to a delay of the depth effect, which resulted in an illogical perspective of the rendered face.
Albeit non of the participants reported this to be an issue, it might be a cause for a the reported Uncanny Valley effect.

\section{Discussion}


The introduction of the design categories answered our research question on \emph{``How to design the communication of an HMD user's state, naturally communicated by eye-gaze, towards an bystander?''} in terms of identifying design alternatives.
In the following, we discuss the results of the user study that explored the category \emph{Visualization}.

\subsubsection{A. Information on the Front of an HMD has a Benefit}

The presentation of information about an HMD user's state, communicated by gaze when no HMD is present, is able to reduce misunderstandings in mixed-presence collaboration.
However, the participants' subjective perception about the benefits is quite ambiguous.

The \abs provoked significantly (21\%) less erroneous reactions than the \blank condition. 
In contrast, the \eyes condition did not show significantly less errors compared to the \blank condition.
Still, the \eyes condition shows a tendency towards preventing errors.
In particular, about 20\% less errors occurred when the gaze cue for inattentiveness and reference were presented compared to having a \blank.

Regarding the reaction time to a gaze cue, we could not find an advantage of presenting information about the gaze.
We assume that the linearity of the task and the introduction of the shared surface provided enough information.
By this, a common ground was established and the participants reacted accordingly, without taking the HMD user's state into account.
The participants might have guessed the correct behaviour in the \blank condition.
Guessing a correct response would also explain the surprisingly fast reaction time (Figure \ref{fig:reactiontime}, \emph{Asking for Assistance}) and high error rate (52\%), when there was nothing displayed on the HMD. 

Interestingly, these drawbacks in performance are not affecting the subjective assessment by the participants.
In particular, the \blank condition was scored the most effective, with 5 times more first place ratings than the \eyes condition and 1.75 times more than the \abs condition (Figure \ref{fig:subrating}). 
The reason is, according to the interviews, that in the \blank condition one does not have to monitor the HMD user (N=12).
However, the lack of gaze cues and therefore missing information about the state of the HMD user, was an issue for almost half of the participants.
In addition, not having information about the users gaze created a social disconnection from the HMD user (N=11).

\subsubsection{B. Abstraction Improves Performance}

As expected the presentation of textual information supports the Non-HMD user in the collaboration as the error rate drops significantly in comparison to the conditions \blank (+ 21\%) and \eyes (+ 13.5\%).
Particular interesting is the error rate in the assist condition with around 40\% more mistakes when interpreting the \blank and \eyes condition compared to the \emph{Abstract\_Visualiza\-tion} condition.
Additionally, the error rate of the \emph{Abstract\_Visualiza\-tion} condition does not show extreme variances between the gaze cues. 
The equal distribution indicates stability for other gaze cues used in human collaboration.

In total, the abstract gaze cue was recognized in 16\% more cases than in the \eyes condition. 
In the \emph{Abstract\_Visua\-lization} condition, the attention of the bystander is drawn to the HMD because of the continuous content change on the display, leading to a flickering effect. This effect is smaller in the \eyes condition, as the displayed content does not change continuously in the idle state. Furthermore, the flickering effect is completely absent in the \blank condition. The result is a lower situation awareness of the Non-HMD user in the \eyes and \blank condition.

We could find a tendency of lower recognition times in the \abs condition (Figure \ref{fig:rectime}).
The time to recognize information on the HMDs display did not show any significance between the two visualization conditions.
Only \emph{referencing to an object} was detected slightly faster, as the action might be dominated by the accompanying head movement.

\subsubsection{C. Abstraction Makes Collaboration Robotic}

The introduction of textual visualization of the user's state was reported as being robotic.
This is caused by processing instructions presented on the screen, leading to a feeling of collaborating with a machine (N=5).

Although the textual visualization shows many indications that it makes collaboration more efficient, 50\% of the participants perceive reading the gaze cues as being disruptive.
The reported interruption is in contrast to not finding an effect on the performance measures and the positive result in the forced choice question on the efficiency of the conditions.
We conclude that the front-facing display should present an even more abstract cue, e.g., Icons or color.
The offside of this approach would be the need to learn the color code's meanings.

\subsubsection{D. Realism does Not Improve Social Presence (Yet)}
Our results indicate that there is a positive effect on social presence when providing the illusion of looking through the HMD.
The presentation of the eyes generated a vital connection for the Non-HMD to the HMD user for about a third of the participants.
It needs to be noted that another third of the participants reported an experience close to an Uncanny Valley effect \cite{Mori12}.
We argue that advances in technology might overcome these effects of the Uncanny Valley.
However a visualization of something that is not possible, looking through the HMD, might even reinforce the feeling of uneasiness.
Designers should keep this in mind and choose a more abstract visualization if any signs of the Uncanny Valley get apparent.

\section{Conclusion}
In this work, we addressed the challenges of failing interaction and the strong social barrier between a Non-HMD and an HMD user.
Based on related work we proposed design categories for information visualization of an HMD user's state towards Non-HMD users.
These categories helped us to identify gaps in the design of information displays for co-located mixed-presence collaboration.

In the user study, we found that presenting information on a front facing display helps to improve collaboration, in particular when abstractly visualizing the users state, in our case as text.
The abstract visualization, therefore, might be beneficial for commercial use cases, e.g., to support a salesperson in a shop taking care of an HMD using costumer.

However, the social barrier due to the covering of the HMD user's face still exists when adding an abstract information visualization, compared to not having any information about the HMD user's state presented at all. 
In particular, Non-HMD users reported that they only followed the instructions that were displayed as text on the screen, instead of working with the HMD user.

To overcome this, we used the idea of creating the illusion of looking through the HMD onto the wearers face.
While the display should continue to be an information display representing the state of the user, the design should create a higher social presence of the HMD user for the Non-HMD user.
Related work showed that the presentation of the real face behind the HMD in a video has a very positive effect.
We could find that for about a third of the Non-HMD users with our front facing display attached to the HMD.
Another third suffered from uncanny valley effects, seeing a face that is not truly realistic.
The difference might be that previous work used remote screens which is similar to watching a movie with special effects.
We don't know if improved technology and the Non-HMD users getting used to seeing a face through the HMD will enable more users to create a social connection.

Designers that address the challenge of social exclusion of HMD users in everyday scenarios should keep in minds these possible tradeoffs.
The way we present information might improve the collaborative performance and therefore lead to longer usage of VR HMDs and a better experience. However, with a wrong design, we might unintentionally amplify social exclusion of the HMD user.

\section{Future Work}
The introduction of categories to design for co-located mixed presence collaboration was explicitly aimed at the communication of the HMD user personal information.
Other work already showed how to present information about the virtual environment and the virtual character the HMD user is controlling. 
Future work might use the design categories of our work to foster designs that support information presentation towards the Non-HMD user while maintaining control about the personal and virtual origin of the information.

Further, the design categories presented in this work can be used as a base for a holistic approach.
It would be valuable to work on a design space that covers all dimensions of co-located mixed-presence collaboration that includes communication in both directions.
Our ultimate goal is to make HMD socially acceptable and enable integration into our everyday life without disturbing our natural human interactions.

\bibliographystyle{ACM-Reference-Format}
\bibliography{ms}

\end{document}